\documentclass[a4paper,fleqn,usenatbib]{mnras}

\usepackage{newtxtext,newtxmath}

\usepackage[T1]{fontenc}
\usepackage{ae,aecompl}

\usepackage{graphicx}	
\usepackage{amsmath}	
\usepackage{amssymb}	
\usepackage{multirow}
\usepackage{longtable}
\usepackage{supertabular,booktabs}
\usepackage{float}
\usepackage{ragged2e}

\newcommand{\Msun}{\ifmmode {$M$_{\odot}}\else{M$_{\odot}$}\fi}

\title[X-rays reveal candidate AGN as SNRs]{X-ray spectroscopy of the candidate AGN in Henize 2-10 and NGC 4178: Likely supernova remnants}

\author[Hebbar et al.]{
Pavan R. Hebbar$^{1}$\thanks{E-mail: hebbar@ualberta.ca},
Craig O. Heinke$^{1}$,
Gregory R. Sivakoff$^{1}$,
Aarran W. Shaw$^{1}$
\\
$^{1}$Department of Physics, CCIS 4-183, University of Alberta, Edmonton, AB, T6G 2E1, Canada
}

\date{Accepted XXX. Received YYY; in original form ZZZ}

\pubyear{2018}

\begin{document}
\label{firstpage}
\pagerange{\pageref{firstpage}--\pageref{lastpage}}
\maketitle

\begin{abstract}
Black holes in dwarf/bulgeless galaxies play a crucial role in studying the co-evolution of galaxies and their central black holes. Identifying massive black holes in dwarf galaxies suggests that the growth of black holes could precede that of galaxies. However, some of the most intriguing candidate active galactic nuclei (AGN) in small galaxies have such low luminosities that the sample is vulnerable to contamination by other sources, such as supernova remnants. We re-analysed {\it Chandra X-ray Observatory} observations of candidate AGN in Henize 2-10 and NGC 4178, considering the potential signals of emission lines in the minimally-binned X-ray spectra.
We find that hot plasma models, which are typical of supernova remnants, explain the observed spectra much better than simple power-law models, which are appropriate for AGN. We identify clear signals of X-ray lines in the faint X-ray source identified with the radio source in Henize 2-10 by Reines et al. 2016. Combining our work with the MUSE measurement of the ionization parameter in this region by Cresci et al. 2017 indicates that this radio and X-ray source is more likely a supernova remnant than an AGN. A similar analysis of the low-count X-ray spectrum of a candidate AGN in NGC 4178 shows that a hot plasma model is about seventeen times more probable than a simple power-law model. Our results indicate that investigation of X-ray spectra, even in a low-count regime, can be a crucial tool to identify thermally-dominated supernova remnants among AGN candidates. 
\end{abstract}

\begin{keywords}
galaxies: Henize~2-10 -- galaxies: NGC~4178 -- galaxies:active -- X-rays:galaxies -- galaxies:starburst -- ISM:supernova remnants
\end{keywords}



\section{Introduction}
\label{sec:intro}
The correlations between key properties of a galaxy(the mass of the bulge, and the velocity dispersion) and the mass of the supermassive black hole (SMBH) hosted by the galaxy \citep[e.g.][]{magorrian1998, gebhardt2000, marconi2003, jahnke2009, merloni2010} are of high importance in astrophysics. These observational results have motivated theories on how galactic bulges and SMBHs evolve together \citep[see][for reviews]{fabian2012, kormendy2013,heckman14}. These theories suggest that the properties of galaxies are regulated, in part, through the feedback from active galactic nuclei (AGN) in the form of radiation and outflows. Irradiation from AGN can evaporate cool gas, while jets and winds heat the warmer gas. Both methods serve to terminate star-formation, as well as the fuel source for accretion onto the AGN. 

While massive galaxies and bright cluster galaxies show clear evidence of AGN feedback, it has been harder to study AGN activity in low-mass galaxies that lack bulges, or have only pseudobulges  \citep[which are not classical bulges;  e.g., ][]{dashyan2018,booth2013}. And some theories suggest that the observed connection between black hole and galaxy bulge mass might only be a result of repeated mergers \citep[ i.e., they need not be physically coupled;][]{jahnke2011}. \citet{Volonteri09} points out that the relation of BH masses to galaxy and bulge masses on the low-mass end will provide key evidence as to the origin of massive BHs, as either direct collapse of $\sim 10^4$ \Msun\  clouds versus initial seeds of $\sim 10^3$ \Msun\  BHs from the collapse of Population III (first-generation) stars.  Recent works have been divided as to whether the standard BH-bulge mass relation continues straightforwardly to lower masses \citep{Barth05,Xiao11,Baldassare15} or shows substantial changes \citep{Greene08,Jiang11}. Thus the detections of AGN powered by lower-mass black holes ($10^3$--$10^6$ \Msun, hereafter intermediate-mass black holes or IMBHs, as opposed to super-massive black holes) in low-mass bulgeless or pseudobulge galaxies are crucial for understanding how SMBHs and galaxies grow.

Increasing efforts have been devoted towards finding IMBHs in dwarf and/or bulgeless galaxies over the past 30 years. For instance, the small, bulgeless spiral galaxy NGC 4395 hosts a faint Seyfert I AGN, verified to be powered by a relatively small BH, of mass $(3.6\pm1.1)\times10^5$ \Msun\  \citep{Filippenko89,peterson2005}. The dwarf elliptical galaxy Pox 52 also hosts a weak AGN, with a central BH engine of mass ($3.2\pm1.0)\times10^5$ \Msun\ \citep{Kunth87,Barth04,Thornton08}. A number of optical spectroscopic searches for ``dwarf" Seyfert nuclei have been conducted \citep[e.g.][]{Ho95,Greene04,Greene07,Reines13,Lemons15,Chilingarian18,Liu18}, identifying a large number of candidate IMBHs.  

A number of methodologies have been used to test whether emission line regions at the centres of galaxies are produced by AGN, including: the ratios of emission lines indicating the ionization state (and thus suggesting an origin either in star formation or from AGN activity \citealt{Baldwin81}); the breadth of H lines \citep{Greene05,Reines13}; and the identification of X-rays from the candidate AGN. The Wide-Field Infrared Survey Explorer (WISE) provides mid-infrared colour information, which has been exploited to identify AGN regardless of obscuration  \citep{Stern12,Satyapal14}. \citet{abel2008} proposed using the luminosity of a [Ne V] line in the mid-infrared, which also allows the study of obscured AGN. 

However, young supernova remnants evolving in a high-density medium can also produce emission lines in a relatively high ionization state, and broad H lines \citep[e.g.][]{Filippenko89b,terlevich1992,Baldassare16}. 
\citet{Hainline16} showed that selection using only mid-infrared colours from WISE is vulnerable to confusion from star-forming galaxies.  The formation of Wolf-Rayet and massive O stars in young starbursts can produce extreme ultraviolet flux that could lead to high ionization of Ne \citep{schaerer1999,kewley2001,lutz1998}. Accurate measurement of the BH mass requires time-consuming reverberation mapping \citep[etc.]{peterson1993,peterson2005,peterson2014} or dynamical spectroscopic measurements \citep{du2017,songsheng2018}, which have been done for relatively few candidate IMBH AGN in small galaxies. Thus, identification of X-rays from candidate AGN has become a crucial element of many campaigns to identify IMBH AGN, sometimes being regarded as the crucial piece of proof necessary to verify a candidate AGN \citep[e.g.][]{Greene07,secrest2012,Chilingarian18}. 

The key element of the present work is the realization that moderate-resolution X-ray spectroscopy, even with relatively small numbers of counts, is capable of distinguishing between some hot plasma models,  typical of young thermal supernova remnants (SNRs)\footnote{Not all SNR X-ray spectra are dominated by thermal ejecta; some, like the Crab and SN 1006, are dominated by non-thermal synchrotron emission from pulsar wind nebulae or shocks.}, and the power-law spectra typical of AGN. This opportunity is provided by the very strong emission lines in the metal-enriched thermal ejecta dominating typical SNR spectra, especially the strong Si lines around 1.8 keV. The ability to identify X-ray spectra produced by thermal SNR emission allows us to constrain whether X-ray emission arising from a candidate AGN is more consistent with an AGN or a SNR. In this paper, we study two test cases in detail.

\subsection{Potential AGN in Henize~2-10 and NGC~4178}

Henize 2-10 is a dwarf starburst galaxy without a central bulge, located at a distance of 9 Mpc \citep{Vacca1992}, with a stellar mass of $3.7 \times 10^{9}$\ \Msun\ \citep{reines2011}. \citet{reines2011} identified an X-ray source located in Henize 2-10 with a steep-spectrum radio source. Comparing the X-ray and radio luminosities of this source, \citet{reines2011} argued that the radio and X-ray luminosities {and their ratio} excluded origins other than an accreting black hole. The Fundamental Plane of black hole activity relates black hole mass, X-ray luminosity, and radio luminosity, for black holes accreting in a radiatively inefficient state \citep[e.g.][]{merloni2003}. Using this plane, \citet{reines2011} argued that the observed radio brightness requires a BH with a mass $\sim2\times10^6$ \Msun, as smaller black holes do not generate this radio brightness at the observed $L_X$. The identification of a BH with such a large mass in a dwarf starburst galaxy without a bulge was quite surprising, as it suggested that large BHs could appear first, before galactic bulges grow. Such a discovery would have strong implications for the evolution of galaxies and BHs.

Deeper {\it Chandra} observations of Henize 2-10 \citep{reines2016} resolved the X-ray-source identified as the candidate AGN into two components. One source, coincident with the radio point source, was suggested by \citet{reines2016} to be the AGN.  Another X-ray source, brighter (in some epochs) and spectrally hard, without a radio counterpart, was suggested to be a high-mass X-ray binary. The weaker X-ray source coincident with the radio source had an X-ray luminosity $L_{0.3 - 10 \mathrm{keV}}\sim 10^{38}$ erg s$^{-1}$. \citet{reines2016} fit the X-ray spectrum of the nuclear source with a power-law, and a thermal plasma model (\textsc{apec}), but their spectral fitting did not discriminate between the models. We argue below that their choice of spectral binning lost information, thus preventing clear discrimination between these models. 

Recently, \citet{cresci2017} performed MUSE(Multi Unit Spectroscopic Explorer) integral field spectroscopy on this portion of Henize 2-10. They found that the optical emission line ratios of gas in the central region of Henize 2-10 were consistent with starburst models, with no indication of AGN ionization. They also show that the revised X-ray luminosity, and X-ray/radio flux ratio, of the candidate AGN identified by \citet{reines2016} are consistent with those of young SNRs. 

NGC 4178 is a SB(rs)dm galaxy with a \textsc{HII} nucleus \citep{ho1997} and H$\alpha$ emission similar to a star-forming galaxy \citep{koopmann2004}. This dwarf galaxy \citep[$M_{*} = 1.3 \times 10^{10}$ \Msun;][]{ho1997} is highly inclined ($i \sim 70\deg$) and located at a distance of $16.8$ Mpc \citep{tully1984} in the Virgo cluster. \citet{satyapal2009} argued for the existence of an AGN in NGC 4178, based on the high mid-IR luminosity observed in a [Ne V] line ($L_{14.32\mu\mathrm{m}} = 8.23 \times 10^{37}$ erg s$^{-1}$), using the arguments of \citet{abel2008}. \citet{secrest2012} used a {\it Chandra} X-ray observation of NGC 4178 to identify an X-ray source corresponding to the IR source. However,  the observed luminosity of the X-ray source is about five orders of magnitude less than that expected based on the [Ne V] luminosity. They postulate that this could be due to strong absorption in the nucleus. \citet{secrest2013} did not detect any optical signature of an AGN in Gemini observations. Thus, the X-ray emission and strong [Ne V] emission are the only evidence for an AGN in this galaxy. 

In this paper we scrutinise the X-ray spectra of the candidate AGNs in Henize 2-10 and NGC 4178 to test whether these candidate AGN may be better explained as SNRs. In Section~\ref{sec:obs}, we describe how we obtained X-ray spectra of the candidate AGNs in Henize 2-10 and NGC 4178. Section~\ref{sec:results} compares various models fit to the observed X-ray spectra. We demonstrate that minimally-binned spectral analysis can robustly identify strong X-ray spectral lines, which can be helpful in discriminating between AGN and SNRs.

\section{Observations and Data reduction}
\label{sec:obs}

\begin{figure*}
\centering
\includegraphics[width=0.43\textwidth]{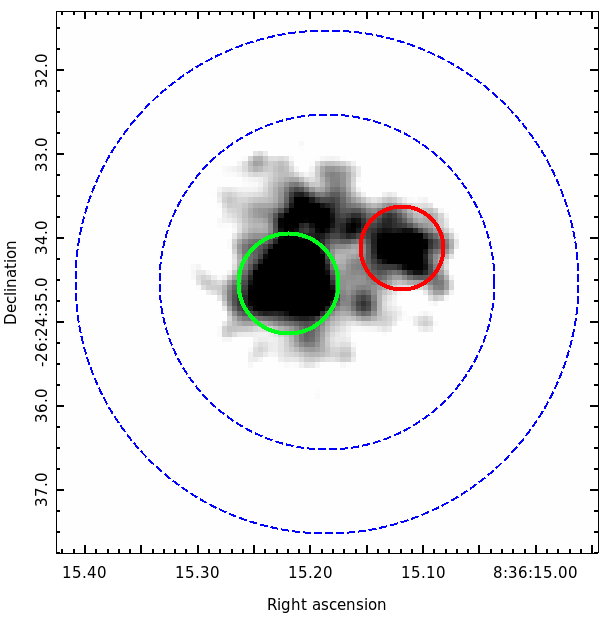} \quad
\includegraphics[width=0.45\textwidth]{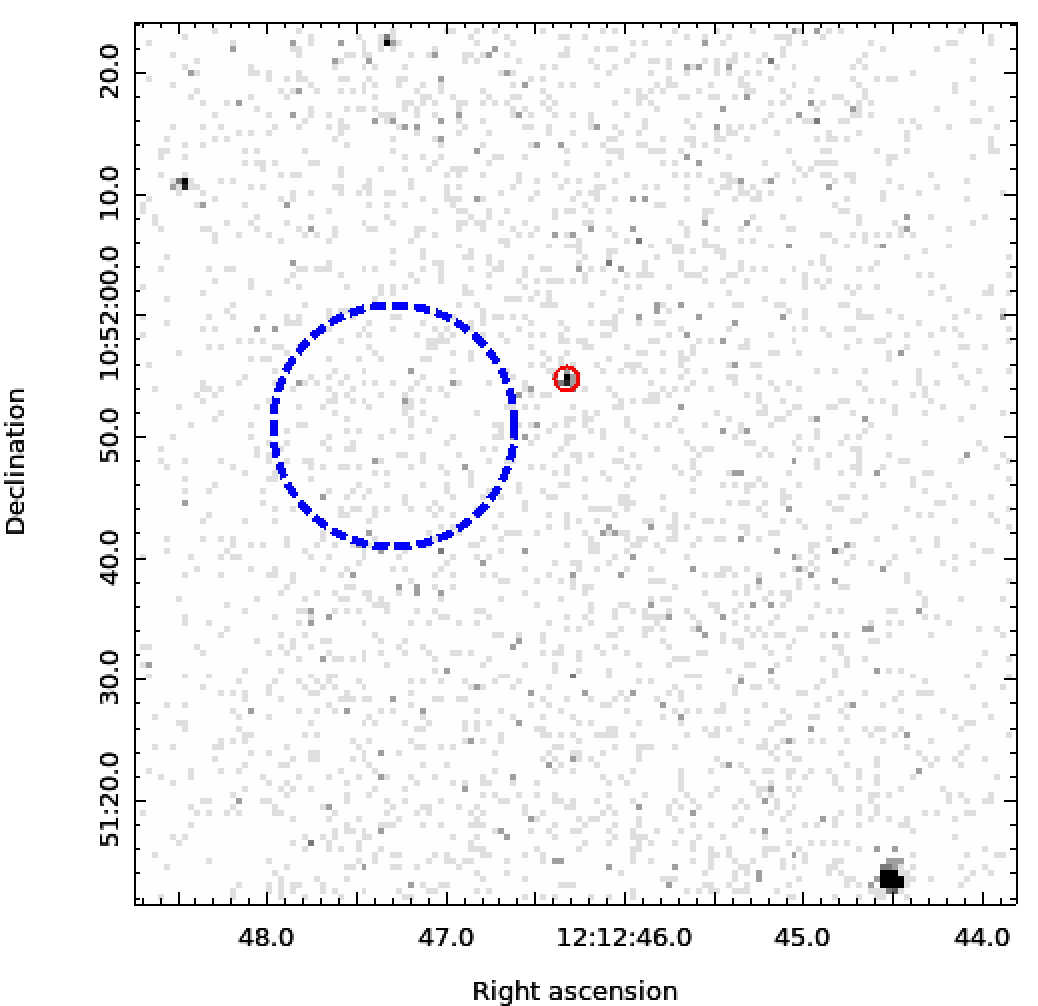}
\caption{({\it Left}) {\it Chandra} 0.3-10 keV X-ray image of the candidate AGN in Henize 2-10, overbinned to $1/8^{th}$ of the ACIS pixel size, and smoothed {with a Gaussian of FWHM= $0\farcs25$}. The $0\farcs5$ source region (red) captures 207 photons. {We also show the bright source to the east of the candidate AGN. The $0\farcs6$ circular region (green) around this source encloses 356 counts.}. The background annular region, shown between two dashed blue circles, contains 598 photons. ({\it Right}) {\it Chandra} 0.3-10 keV image of the candidate AGN in NGC 4178. The source region (red) contains 39 photons. The background circular region (dashed blue circle) contains 189 photons.}
\label{fig:source_bkg}
\end{figure*}

We used the high-resolution {\it Chandra} X-ray observations of Henize 2-10 in February 2015 \citep[ObsID 16068, PI: Reines;][]{reines2016}) and of NGC 4178 in February 2011 \citep[ObsID 12748, PI: Satyapal;][]{secrest2012}) to study these X-ray sources. 
Both observations were taken using the ACIS-S instrument, in the VFAINT mode and FAINT mode, respectively. 

We used \textsc{CIAO} version 4.9 \citep{fruscione2006} for the data reduction and the construction of the X-ray spectra. We reprocessed the initial data according to CALDB 4.7.6 standards using the \texttt{chandra\_repro} command. We set the \texttt{check\_vf\_pha} parameter to ``no'' to prevent the removal of any good events. The default option of the \texttt{pix\_adj} parameter now uses the EDSER algorithm \citep{li2004}, which results in more-precise sub-arcsecond resolution. We constructed X-ray spectra using the \texttt{specextract} command, which produces effective area files corrected for the fraction of the point-spread function extracted in the spectrum. We used  photons with energies between $0.3-10.0$ keV for our analysis.

Because of the low photon counts, grouping these spectra with 15 photons in each bin would provide an inadequate number of bins to constrain different models. Such spectra would also not resolve any narrow features that might exist in the spectrum. Thus, we regrouped the spectra using the \texttt{dmgroup} command such that each group includes at least 1 event. The spectra generated were analysed using \textsc{XSPEC} version 12.9.1m, using the C-statistic \citep{cash1979} modified to deal with  backgrounds, which can handle few to no counts per bin (note that binning to keep 1 count/bin is recommended\footnote{https://heasarc.gsfc.nasa.gov/xanadu/xspec/manual/node304.html}).

\subsection{Henize 2-10}
The central region of Henize 2-10 is extremely compact with multiple X-ray sources within a $2 \arcsec$ region, plus three other clear X-ray sources within a $50 \arcsec$ region. The X-ray image of the central region of Henize 2-10 is dominated by diffuse emission and a highly variable off-nuclear hard X-ray source suggested by \citet{reines2016} to be a high-mass X-ray binary. However, the likely HMXB is not visible in the 160 ks 2015 observation of Henize 2-10 used here. Thus, there is relatively little contamination of the weak X-ray source and candidate AGN, compared to the other X-ray observations of Henize 2-10 (The HMXB, which is  $0\farcs7 $ from the candidate AGN,  is $\sim 10$ times brighter than our source in the remaining observations. Thus we cannot use the ObsIDs 2075 (20 ks) and 16069 (40 ks), taken in 2001 March and on 2015 February 16, respectively, for this analysis.). To clearly resolve the candidate AGN from the nearby diffuse emission and select the source region, we re-binned the image to $1/8^{\rm th}$ of the ACIS-S pixel size and convolved this image with an FWHM$=0 \farcs 25$\, ($\sigma \approx 0.44$ pixels) Gaussian kernel using the \texttt{aconvolve} command (see Fig.~\ref{fig:source_bkg}). We extracted X-ray photons from a $0\farcs5$ circular region centred on this X-ray source ($\alpha = 8^{\mathrm{h}}36^{\mathrm{m}}15 \fs 13$; $\delta = -26^{\circ} 24\arcmin 34 \farcs 08$). This radius was chosen to encompass the maximum source photons while minimising  photons from the diffuse emission around the nuclear source (see Fig.~\ref{fig:source_bkg}, left). We selected background photons from an annulus of radius $2-3\arcsec$, excluding the bright diffuse emission directly east of the AGN candidate. 

\subsection{NGC 4178}
The sub-arcsecond angular resolution of the {\it Chandra X-ray Observatory} clearly resolves the nuclear X-ray source from nearby bright X-ray sources (see Fig.~\ref{fig:source_bkg}, right). We selected a circular region of $1 \arcsec$ around the source ($\alpha = 12^{\mathrm{h}} 12^{\mathrm{m}}46 \fs 32$; $\delta = +10^{\circ} 51\arcmin 54 \farcs 61$) to collect the source photons. We used a $10 \arcsec$ region nearby, lacking bright X-ray sources, for the background.

\section{Results and Discussion}
\label{sec:results}
In each fit, we calculated the best-fitting parameters using the C-statistic, which is the maximum-likelihood-based statistic for Poisson data \citep{cash1979}, as the fit statistic (e.g. ``statistic cstat" in XSPEC), with correction for background subtraction, aka the ``W-statistic"\footnote{https://heasarc.gsfc.nasa.gov/xanadu/xspec/manual/node304.html}.
To evaluate  the relative quality of different models, we used the Akaike Information Criterion values \citep{akaike1974}, with correction for small sample sizes \citep[AICc; see][]{cavanaugh1997}:
\begin{equation}
\mathrm{AICc} = 2k + \mathrm{cstat}_{min} + \frac{2k^2 + 2k}{n-k-1},
\end{equation}
where $k$ is the number of parameters and ${\rm cstat}_{\rm min}$ is the negative of twice the logarithm of the likelihood of the model, with best-fitting parameters assuming  Poisson statistics. A smaller AICc value indicates a better likelihood. The quantity $\exp$((AICc$_i$ - AICc$_j$)/2) gives the relative likelihood of model $j$ with respect to model $i$. A small value (e.g. 1/20) of this quantity indicates that model $i$ is preferred, by the inverse of this factor (e.g. by a factor of 20) over model $j$. 

We also use XSPEC's ``goodness" simulations to calculate the quality of individual fits. These simulations use the Cramer-von Mises (CvM) statistic \citep{cramer1928,vonMises1928} (e.g. we set ``statistic test cvm" in XSPEC). They measure the fraction of simulations of the model with a CvM statistic smaller (better) than the CvM statistic of the data; a fraction close to 1 (e.g. 99\%) indicates the model can be rejected with this level (e.g. 99\%) of confidence.

For both sources, we chose a simple absorbed power law spectrum for an initial fit, and when that proved inadequate to explain the emission features (based on the quality of the fit), moved on to use hot thermal plasma models . For modelling photoelectric absorption, we used the \textsc{tbabs} model with typical interstellar ({\it wilm}) abundances \citep{wilms2000}. For both sources, we find that the hot plasma models explain the observed spectra better than a simple power law. The results of our spectral fitting are summarized in Tables~\ref{table:spec_hen2-10} \& \ref{table:spec_ngc4178}. For easy visualization, the plotted spectra of Henize 2-10 and NGC 4178 in this paper were re-binned such that each bin had four and three counts, respectively. We also tested the effects of modelling the background for the Henize 2-10 candidate AGN, rather than subtracting it. As the background makes only a small ($\sim10$\%) contribution to the source flux, we found negligible differences in the fits that we tested, with all fitted parameters lying within the errors found with background subtraction.

\subsection{Henize 2-10}
\begin{figure}
\centering
\includegraphics[width=\columnwidth]{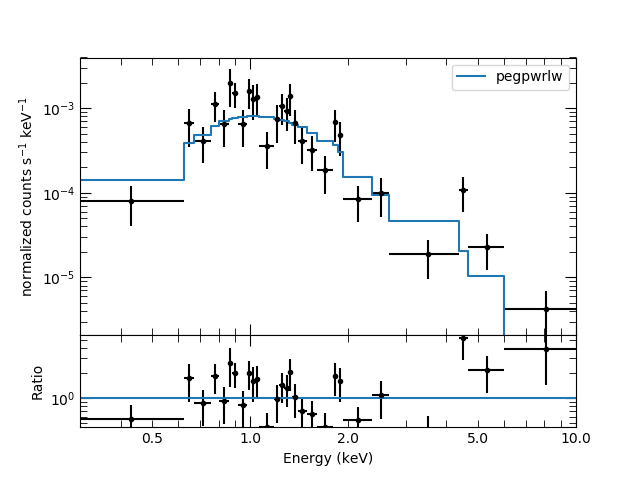}
\caption{Power-law fit to the X-ray spectrum of the candidate AGN in Henize 2-10, with $\Gamma = 2.7_{-0.5}^{+0.5}$ and $N_H = 2.3_{-1.4}^{1.7} \times 10^{21}$ cm$^{-2}$. The C-statistic is 131.5 for 3 free parameters (93 d.o.f). Though the fit can explain the overall shape of the spectrum, it fails to model the observed narrow features.}
\label{fig:hen_pegpw}
\end{figure}

From Fig.~\ref{fig:source_bkg}, we see that there is another bright X-ray source to the east of our candidate AGN, even in ObsID 16068. However, the convolved image shows that the X-ray emission from the two sources are distinct and clearly resolved by {\it Chandra}. We analyse the X-ray flux from this eastern source in later sections. We binned the $\sim 200$ photons observed from the candidate AGN into $\approx 96$ bins, each with one or more count per PHA (energy) bin. The minimum $N_H$ was fixed to $9.1 \times 10^{20}$ cm$^{-2}$ to account for Galactic absorption in the direction of Henize 2-10 \citep{kalberla2005}. Fitting an absorbed power law (\textsc{tbabs*pegpwrlw}) to the spectrum gives $\Gamma = 2.7_{-0.5}^{+0.5}$ and $N_H = 2.3_{-1.4}^{+1.7} \times 10^{21}$ cm$^{-2}$, consistent with the spectral fit found by \citet{reines2016}. This fit is shown in Fig.~\ref{fig:hen_pegpw}, achieving a C-statistic of 131.47. We created  $10^5$ realizations of the power-law spectrum using the \texttt{goodness} command, and found that $\sim 99\%$ of them have a lower CvM statistic than the data. This indicates that a simple power-law is a poor description of the observed spectrum. 

In Fig.~\ref{fig:hen_pegpw}, we can see that the power law model predicts the overall trend in counts versus energy, but fails to explain several narrow features clearly visible by eye in the observed spectrum. On studying the residuals, we see that the higher counts at $\approx 1.3$ keV and $\approx 1.9$ keV align with the K$\alpha$ emission lines of Mg and Si. The overall shape of the spectrum, with evidence of strong lines near 1.0, 1.3, 1.9, and (perhaps) 2.5 keV is reminiscent of the spectral shapes of supernova remnants, with lines around those energies due to Ne and/or Fe, Mg, Si, and S, respectively (e.g. Cas A SNR: \citealt{Holt94,Hughes00}; Tycho SNR: \citealt{Badenes06}; SN 1987A: \citealt{Michael02}).There also appears to be a feature at $\sim 4.5$ keV, which we discuss below.

\begin{figure}
\centering
\includegraphics[width=\columnwidth]{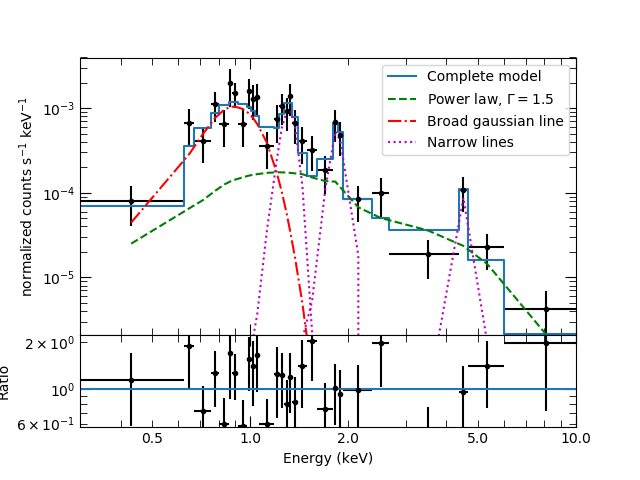}
\caption{X-ray spectrum of the AGN candidate in Henize 2-10, here modelled with 4 Gaussian emission lines added to a power-law continuum. The broad line at 0.8 keV may be due to the K$\alpha$ lines of O and Ne, along with Fe-L lines. The narrow lines at 1.3 \& 1.8 keV correspond to K$\alpha$ lines of Mg and Si, respectively. The nature of the narrow 4.5 keV line is unclear. This figure shows strong evidence for multiple emission lines in this X-ray spectrum.}
\label{fig:hen_pegpw_gggg}
\end{figure}

As a first step, we model the spectrum with an absorbed power-law plus Gaussians at the locations of strong residuals (Fig.~\ref{fig:hen_pegpw_gggg}). We find that four lines are strongly required. Narrow ($\sigma<0.1$) lines are required at $1.85_{-0.05}^{+0.06}$ keV (which we associate with Si XIII K$\alpha$, lab energy 1.83 keV\footnote{http://www.atomdb.org/Webguide/webguide.php}), $1.31_{-0.04}^{+0.02}$ keV (which we associate with Mg XI K$\alpha$, 1.33 keV), and $4.56\pm0.11$ keV, which does not have an immediately obvious nature. A broad ($\sigma=0.18_{-0.06}^{+0.09}$ keV) emission feature is also required at 0.78$^{+0.06}_{-0.22}$ keV, which may be explained by a combination of O VIII (0.65 keV), Ne IX K$\alpha$ (0.91 keV), and/or Fe L-shell (covering roughly 0.8--1.1 keV). When we add these four Gaussians to the power-law model,  the C-statistic drops by $\sim 45$ with 10 fewer degrees of freedom. Comparing the AICc statistics (Table \ref{table:spec_hen2-10}) for the power-law fit, vs. the fit including four Gaussians, indicates that the 4-Gaussian fit is preferred by a factor of $\sim$37,000.

These strong emission lines suggest thermal plasma at low temperatures and/or high abundances, consistent with thermally emitting ejecta from a supernova remnant. We therefore  attempt to fit the spectrum of the Henize 2-10 AGN candidate with self-consistent physically motivated thermal plasma models, beginning with a simple collisionally ionized model.

The \textsc{apec} model has a single parameter for all metal abundances. Since we do not see a strong Fe K emission line in our spectrum (e.g. at 6.7 keV), the single \textsc{apec} model fails to fit the observed emission lines of Mg and Si. Thus we employed the \textsc{vapec} model that allows fits to individual relative abundances of metals. We allowed the abundances of O, Ne, Mg, Si and S to vary relative to solar values, since these are the abundant elements with strong lines in the observable spectral range.
Fitting an absorbed collisionally ionized plasma emission model (\textsc{tbabs*vapec}) gives $kT = 2.5_{-0.8}^{+0.9}$ keV and 
prefers high abundances of O ($176_{-169}^{+824}$) and Ne ($190_{-175}^{+810}$), with Si ($5_{-5}^{+995}$), S ($7_{-7}^{+228}$), and Mg ($0^{+249}_{-0}$) abundances basically unconstrained. Since the errors on the relative abundances of Mg, Si and S are large, and consistent with solar abundances, we fix the abundance of these elements to solar (1). Fitting the X-ray spectrum with this constraint gives $N_H = 9_{-0}^{+15} \times 10^{20}$ cm$^{-2}$, $kT = 2.5_{-0.9}^{+0.9}$ keV, and super-solar abundances of O and Ne ($\sim 1000$). \footnote{The lower limit on $N_H$ reaches our hard limit, while the abundances of O \& Ne reach the hard upper limit.} The spectrum matches the observed features with K$\alpha$ emission lines of O and Ne. Despite having two additional parameters in comparison to an absorbed power law model, the C-statistic decreases only marginally (by $6.47$). Thus, though this fit is better than a simple absorbed power-law,  it is not a statistically significant improvement over the power-law model. We also performed ``goodness" simulations using $10^5$ realizations of the model to test the quality of the fit. Due to the large error-bars in the abundance parameters, we used the ``\texttt{nosim}'' option in the \texttt{goodness} command in XSPEC for our simulations. Similar to the absorbed power-law spectra, we find that $\sim 99\%$ of the realizations have a lower CvM statistic confirming that a single temperature hot plasma model is a poor fit to the observed spectrum. Also, the high relative abundance of O and Ne with respect to Mg and Si seems unphysical.

\begin{figure}
\centering
\includegraphics[width=\columnwidth]{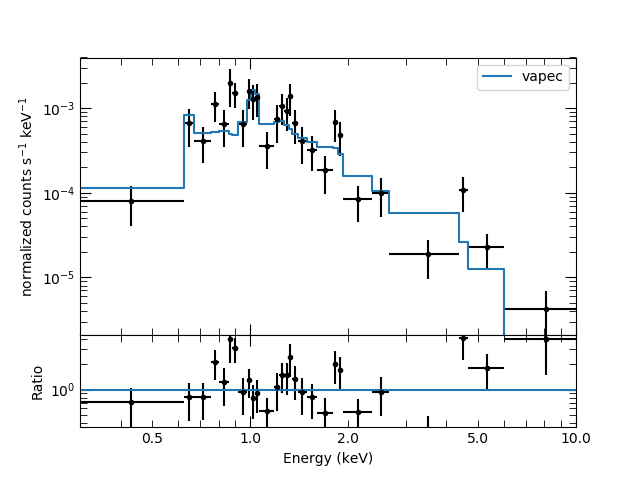}
\caption{X-ray spectrum of the AGN candidate in Henize 2-10, here modelled by an absorbed collisionally ionized plasma emission model with a single temperature, $kT = 2.5_{-0.9}^{+0.9}$ keV. The model fits the emission features of O and Ne well, but fails to fit  other narrow spectral features that are suggestive of emission from Mg, Si, and S.} 
\label{fig:hen_vapec}
\end{figure}

The low abundances of Mg \& Si in this fit seem unlikely, since we see evidence for apparent emission lines near the K$\alpha$ energies of Mg XI and Si XIII. Adding Gaussian emission lines, with $\sigma = 0.01$ keV, at energies 1.31 \& 1.85 keV  decreases the C-statistic by $\sim 14$, while decreasing the degrees of freedom by 4. The change in the AICc values indicates that adding these Gaussian lines increases the likelihood by a factor of $\sim 20$.  The data thus suggest that these lines are real, but that the single-temperature plasma model does not get their energies correct.

Another way of showing the statistical significance of these lines is by looking at the probability of obtaining the observed number of counts in these energy ranges, given the spectral fit to the remainder of the spectrum. While the single temperature hot plasma model predicts a total of 26.4 photons in the energy ranges 1.2-1.4  and 1.75 - 1.95 keV, the data show 46 photons in these bins. Assuming a Poissonian distribution, this has a single-trial probability of  0.2\%.

The single-temperature model predicts the Mg K$\alpha$ line to be at slightly higher energies ($\sim 1.47$ keV) than observed (1.31 keV). 
At the required temperature to explain the high-energy photons as bremsstrahlung, the Mg will be principally in Mg XII, which has a K$\alpha$ line at 1.47 keV, rather than the observed 1.31 keV better fit by Mg XI. Thus, a lower temperature plasma could match the observed Mg emission line, but would not explain the observed continuum emission at higher energies. We consider three models to address this problem.

\begin{figure}
\centering
\includegraphics[width=\columnwidth]{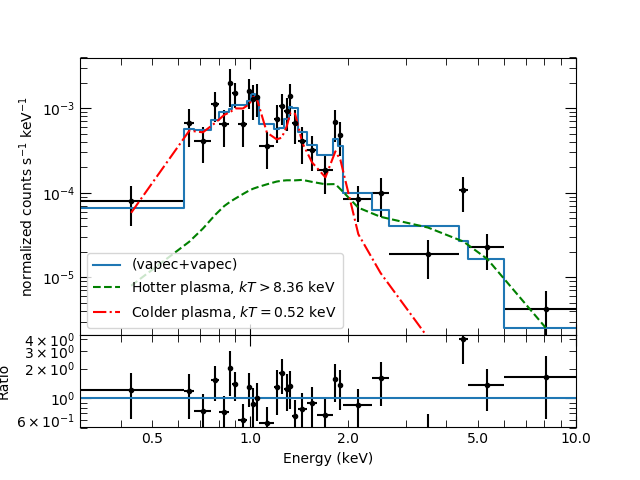}
\caption{X-ray spectrum of the AGN candidate in Henize 2-10, here modelled by an absorbed collisionally ionized plasma model with two different temperatures, $kT > 8.4$ keV \& $kT = 0.5_{-0.1}^{+0.3}$ keV . We find super-solar abundances of O, Ne, Mg, Si ($=11_{-7}^{+24}$) . 
This model captures the continuum emission as well as all the emission features, except the one at $\sim 4.5$keV.}
\label{fig:hen_vapec2}
\end{figure}

First, we consider a two-temperature plasma to model the observed spectrum (\textsc{tbabs*(vapec+vapec)}). Similar models have often been used to model SNR emission, and have been suggested to indicate cooler emission from the reverse shock and hotter emission from the blast wave \citep[e.g.][]{Jansen88, Willingale02}. To minimise the number of free parameters, we only allowed the abundances of the (measurable) elements O, Ne, Mg \& Si to vary. When fitting this model, we see that the error bars in the abundances of these four elements overlap. Numerical simulations of supernovae with progenitor masses from 9 to 120 \Msun\ in \citet{sukhbold2016} show that the relative abundance ratios of these elements with respect to Si are $\sim O(1)$. Since the errors on the abundance values are large in our case, we linked the abundances of O, Ne, Mg and Si in both plasmas. This model gives $N_H = 2.7_{-1.8}^{+2.8} \times 10^{21}$cm$^{-2}$, $kT = 0.52_{-0.11}^{+0.24}$ for the cold plasma and $kT \geq 8.36$ keV for the hot plasma (see Fig.~\ref{fig:hen_vapec2}). We find the relative abundances of the linked elements O, Ne, Mg, Si and S to be $11_{-7}^{+24}$ times solar. This model is shown in Fig.~\ref{fig:hen_vapec2}. The Fe abundance prefers to be smaller than the abundances of O, Ne, Mg, Si \& S. Fixing the abundances of O, Ne, Mg, \& Si to 11.0, we constrain the Fe abundance to be $1.0_{-0.85}^{+1.08}$. Compared to the absorbed power-law model, the C-statistic decreases by 31.82 while the degrees of freedom only decreases by 5. Thus, the AICc value decreases by 24.84, implying that this model is $\sim 2.5\times 10^{5}$ times more likely. The ``goodness'' simulations show that $\sim 59\%$ of the realizations have a lower CvM statistic that the observed data, showing that this model is a good fit to the observed data.

The lower ionization state of Mg \& Si could also be explained by a plasma that is not yet in ionization equilibrium, as is often the case in young SNRs. A \texttt{vnei} model with a lower ionization time-scale \citep[The ionization time-scale to attain equilibrium at $10^7$ K  is $\sim 10^{11}-10^{12}$ s cm$^{-3}$ for Mg and Si;][]{smith2010} can explain the continuum emission at higher energies, while explaining the low-ionization emission lines of Mg and Si. The relative abundances of O, Ne, Mg \& Si were linked together as in the previous models. When fitting this model, we find $N_H = 5.9_{-2.2}^{+1.8} \times 10^{21}$ cm$^{-2}$, $kT > 22$ keV and ionization time-scale $\tau = 1.6_{-0.7}^{+1.1}\times 10^{10}$ s cm$^{-3}$. The ionization time-scale is similar to that of the young supernova remnant Cassiopeia A \citep{Willingale02} .  Though we find the abundance of O, Ne, Mg and Si to be consistent with solar values within their error limits ($1.7_{-0.7}^{+1.2}$), the low ionization time-scale of this model argues for the X-rays being generated by a SNR, rather than an AGN. This model has similar AICc and CvM statistic values as the two-temperature plasma model, suggesting that both models are equally likely.

The emission at higher energies could also be due to synchrotron emission, e.g. from a pulsar wind nebula or synchrotron-dominated shocks, instead of a hotter plasma component. Synchrotron spectra can typically be  modelled using a simple power-law model, and many SNRs show such a strong power-law component. Thus we also try fitting a \textsc{tbabs*(pegpw + vapec)} model to the observed spectrum. Though the best-fitting model does indicate super-solar abundances, the power-law component is very hard($\Gamma = 0.4_{-1.3}^{+1.1}$). Using $\Gamma = 2.7$ \citep[typical of forward shocks in SNRs; eg.][]{torii2006,tamagawa2009}, gives a poorer fit (the ``goodness" exceeds 95\%, indicating a poor fit). Thus, synchrotron emission from shocks in a SNR does not appear to explain the flux at higher energies well.

The observed X-ray spectrum also shows an increase in the photon count rate at $2.6$ keV, which is very close to the K$\alpha$ emission line of S XVI (2.62 keV). This can be incorporated into the \texttt{tbabs*(vapec+vapec)} model by allowing for an increased abundance of S in the hot plasma. Linking the abundance of S in both plasmas brings its abundance down, since we do not see any emission corresponding to the expected S XV line that should be produced by the plasma at the lower temperatures. Similarly, a higher ionization time-scale of S in the \texttt{vnei} model($\sim 10^{13}$ s cm$^{-3}$) could also account for the relative over-abundance of S XVI as compared to S XV. Understanding why S behaves differently from the other elements requires further observations.  

\begin{figure}
\centering
\includegraphics[width=\columnwidth]{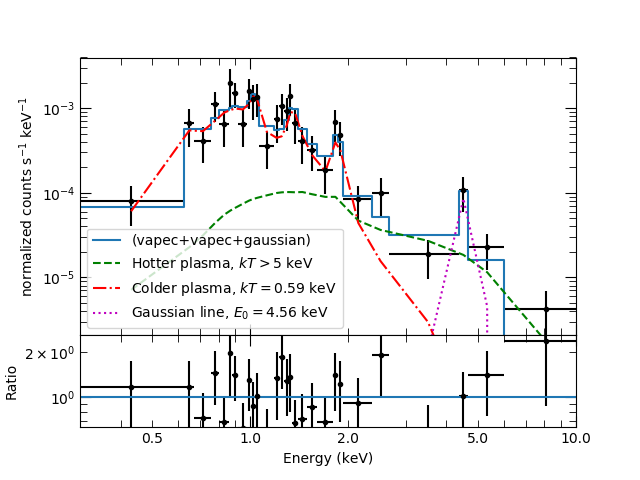}
\caption{X-ray spectrum of the AGN candidate in Henize 2-10, here modelled with a Gaussian emission line at 4.6 keV added to the two-temperature plasma. Since the cstat decreases by $\sim 10$ on addition of the Gaussian, it might indicate a real feature.}
\label{fig:hen_vapec_gauss}
\end{figure}
\begin{table*}
\centering
\caption{Summary of spectral analysis results for the candidate AGN in Henize 2-10.}
\label{table:spec_hen2-10}
\begin{tabular}{c  c  c  c  c  c }
\hline
\multirow{2}{*}[0.0cm]{\bf Model}  &	\multirow{2}{*}[0.0cm]{\bf Parameters} & \multirow{2}{*}[0cm]{\bf Parameter Values} &	\multirow{2}{*}[0cm]{$\mathbf{\Delta}${\bf cstat}$_{\mathbf{pl}}$/$\mathbf{\Delta}${\bf d.o.f}$_{\mathbf{pl}}$} &	\multirow{2}{*}[0cm]{\bf AICc value} &	{\bf log(CvM)} \\
 & & & & & {\bf (goodness)} \\
\hline
\multirow{3}{1in}[-0.1cm]{\textsc{tbabs*pegpwrlw}} &	$N_H$ &	$2.3_{-1.4}^{+1.7} \times 10^{21}$ cm$^{-2}$ &	\multirow{3}{*}[-0.1cm]{$0.0/0$} & 	\multirow{3}{*}[-0.1cm]{$137.73$} &	\multirow{3}{0.5in}[-0.1cm]{\centering $-6.32$ \\$(99 \%)$} \\[0.1cm]
	& $\Gamma$ &	$2.71_{-0.46}^{+0.52}$ &	&	& 	\\[0.1cm]
    & Flux$_{\rm unabsorb}$ & $1.61_{-0.48}^{+1.10} \times 10^{-14}$ ergs cm$^{-2}$ s$^{-1}$ & & & \\[0.1cm] \hline
 \multirow{9}{1in}[-0.3cm]{\textsc{tbabs*(pegpwrlw \\+gauss+gauss \\+gauss+gauss)}} &	$N_H$ &	$1.3_{-0.4}^{+2.8} \times 10^{21}$ cm$^{-2\ *}$ &	\multirow{9}{*}[-0.3cm]{$-45.26/-10$} & 	\multirow{9}{*}[-0.3cm]{$116.70$} &	\multirow{9}{0.5in}[-0.3cm]{\centering $-9.37$ \\$(10 \%)$} \\[0.1cm]
	& $\Gamma$ &	$1.5_{-0.9}^{+0.8}$ &	&	& 	\\[0.1cm]
    & Flux$_{\rm unabsorb}$ & $4.8_{-1.6}^{+3.7} \times 10^{-14}$ ergs cm$^{-2}$ s$^{-1}$ & & & \\[0.1cm]
    & $E_1$ & $0.78_{-0.22}^{+0.06}$ keV &	& & \\[0.1cm]
    & $\sigma_1$ &	$0.18_{-0.06}^{0.09}$ keV & & & \\[0.1cm]
    & $E_2$ & $1.31_{-0.04}^{+0.02}$ keV & & & \\[0.1cm]
    & $E_3$ & $1.85_{-0.05}^{+0.06}$ keV & & & \\[0.1cm]
    & $E_4$ & $4.56 \pm 0.11$ keV & & & \\[0.1cm]
    & $\sigma_2,\sigma_3, \sigma_4$ &	$0.04_{-0.04}^{+0.06}$ keV & & & \\[0.1cm] \hline
\multirow{5}{1in}{\vspace{-0.3cm} \textsc{tbabs*vapec}} & $N_H$ &	$9.1_{-0.0}^{+15.0} \times 10^{20}$ cm$^{-2\ *}$ &		\multirow{5}{*}{\vspace{-0.3cm} $-6.47/-2$} & \multirow{5}{*}{\vspace{-0.3cm} $135.67$} &	\multirow{5}{0.5in}[-0.1cm]{\centering  $-6.40$ \\$(99\%)$} \\[0.1cm]
	& $kT$	& $2.45_{-0.83}^{+0.91}$ keV&	&	&	\\[0.1cm]
	& O &	$960_{-930}^{+40\ **}$ & & & \\[0.1cm]
    & Ne &	$970_{-940}^{+30\ **}$ & & & \\[0.1cm]
    & E.M.$^{a}$   & $1.39_{-0.31}^{+25.4} \times 10^{59}$ cm$^{-3}$& & & \\[0.1cm] \hline
\multirow{6}{1in}{\vspace{-0.5cm} \textsc{tbabs*(vapec+vapec)}} & $N_H$ &	$2.7_{-1.8}^{+2.8} \times 10^{21}$ cm$^{-2\ *}$ &		\multirow{6}{*}{\vspace{-0.5cm} $-31.82/-3$} & \multirow{6}{*}{\vspace{-0.5cm} $112.59$} &	\multirow{6}{0.5in}[-0.25cm]{\centering $-8.12$\\$(59\%)$} \\[0.1cm]
	& $kT_{\rm hot}$ & $> 8.36$ keV &	 & & \\[0.1cm]
    & $kT_{\rm cool}$ &	$0.52_{-0.11}^{+0.24}$ keV & & &\\[0.1cm]
	& O, Ne, Mg, Si	&	$11_{-7}^{+24}$ &	&	&	\\[0.1cm]
    & E.M.$_{\rm hot}^{a}$ & $3.3_{-2.2}^{+1.83} \times 10^{60}$ cm$^{-3}$ & & & \\[0.1cm]
    & E.M.$_{\rm cool}^{a}$ & $1.72_{-0.96}^{+5.66} \times 10^{60}$ cm$^{-3}$ & & &\\[0.1cm]
\hline
\multirow{5}{1in}{\vspace{-0.5cm} \textsc{tbabs*vnei}} & $N_H$ &	$5.9_{-2.2}^{+1.8} \times 10^{21}$ cm$^{-2}$ &		\multirow{5}{*}{\vspace{-0.5cm} $-29.31/-2$} & \multirow{5}{*}{\vspace{-0.5cm} $112.83$} &	\multirow{5}{0.5in}[-0.2cm]{ $-8.15$\\$(38\%)$} \\[0.1cm]
	& $kT$ & $>22$ keV & & & \\[0.1cm]
    & O, Ne, Mg, Si & $1.7_{-0.7}^{+1.2}$ & & & \\[0.1cm]
    & $\tau$ & $1.6_{-0.7}^{+1.1} \times 10^{10}$ s cm$^{-3}$ & & & \\[0.1cm]
    & E.M$^{a}$ & $3.9_{-1.4}^{+2.0} \times 10^{60}$ cm$^{-3}$ & & & \\[0.1cm] \hline
\end{tabular}\\
\justify
Note: For models with \textsc{vapec} \& \textsc{vnei}, the abundances of all the elements that were not specified in the table were fixed to solar values (1). The unabsorbed flux is calculated in the $0.3-10$ keV interval. The ``goodness" values, below the CvM statistic values, denote the percentage of realizations of the model that have a lower CvM statistic than the data.
Thus, a large ``goodness" value (e.g. 99\%) indicates a small (e.g. 1\%) probability of obtaining such a poor fit by chance--and thus a high probability that the model is not the correct description of the data.
Comparing the $\Delta$cstat/$\Delta$d.o.f and the AICc values between different models, we see that the Henize 2-10 AGN candidate is more likely to be a SNR than an AGN. The values of CvM statistic and ``goodness" further confirm our results. \newline
$^{*}$ Lower error bound reaches the lower hard limit. \newline
$^{**}$ Upper error bound exceeds the upper hard limit. \newline
$^{a}$ E.M stands for Emission Measure = $\int n_e n_H dV$
\end{table*}

\begin{table*}
\centering
\contcaption{Summary of spectral analysis results for Henize 2-10: models with a Gaussian line added at $\sim 4.6$ keV}
\label{table:spec_hen2-10cont}
\begin{tabular}{c  c  c  c  c  c }
\hline
\multirow{2}{*}[0.0cm]{\bf Model}  &	\multirow{2}{*}[0.0cm]{\bf Parameters} & \multirow{2}{*}[0cm]{\bf Parameter Values} &	\multirow{2}{*}[0cm]{$\mathbf{\Delta}${\bf cstat}$_{\mathbf{pl}}$/$\mathbf{\Delta}${\bf d.o.f}$_{\mathbf{pl}}$} &	\multirow{2}{*}[0cm]{\bf AICc value} &	{\bf log(CvM)} \\
 & & & & & {\bf (goodness)} \\
\hline
\multirow{8}{1in}{\vspace{-0.7cm} \textsc{tbabs*(vapec+vapec \\+gauss)}} & $N_H$ &	$2.2_{-1.3}^{+1.9} \times 10^{21}$ cm$^{-2\ *}$ &		\multirow{8}{*}{\vspace{-0.7cm} $-40.69/-5$} & \multirow{8}{*}{\vspace{-0.7cm} $108.44$} &	\multirow{8}{0.5in}[-0.3cm]{\centering $-8.29$ \\$(46 \%)$} \\[0.1cm]
	&  $kT_{\rm hot}$ &	> 5 keV & & & \\[0.1cm]
    & $kT_{\rm cool}$ & $0.59_{-0.16}^{+0.20}$ keV & & & \\[0.1cm]
    &  O, Ne, Mg, Si & $11_{-7}^{+46}$ & & & \\[0.1cm]
    & E.M.$_{\rm hot}^{a}$ & $2.3_{-1.3}^{+1.7} \times 10^{60}$ cm$^{-3}$& & & \\[0.1cm]
    & E.M.$_{\rm cool}^{a}$ & $1.6_{-0.7}^{3.4} \times 10^{60}$ cm$^{-3}$ & & & \\[0.1cm]
    & $E_0$ & $4.56 \pm 0.11$ keV & & & \\[0.1cm]
    & $\sigma$ & 0.1 keV & & & \\[0.1cm]
    & Flux$_{\rm unabsorb, gauss}$ & $6.6_{-4.4}^{+6.6} \times 10^{-16}$ ergs cm$^{-2}$ s$^{-1}$ & & & \\[0.1cm] \hline
\multirow{7}{*}{\vspace{-0.5cm} \textsc{tbabs*(vnei+gauss)}} & $N_H$ &	$4.9_{-2.0}^{+2.1} \times 10^{21}$ cm$^{-2}$ &		\multirow{7}{*}{\vspace{-0.5cm} $-39.40/-4$} & \multirow{7}{*}{\vspace{-0.5cm} $107.34$} &	\multirow{7}{0.5in}[-0.3cm]{\centering $-8.50$\\$(22\%)$} \\[0.1cm]
	& $kT$ & $> 16$ keV & & & \\[0.1cm]
    & O, Ne, Mg, Si & $2.3_{-1.0}^{+2.0}$ & & & \\[0.1cm]
    & $\tau$ & $2.1_{-0.9}^{+1.0} \times 10^{10}$ s cm$^{-3}$ & & & \\[0.1cm]
    & E.M.$^{a}$ & $2.8_{-1.1}^{+1.6} \times 10^{60}$ cm$^{-3}$ & & & \\[0.1cm]
    & $E_0$ & $4.56 \pm 0.11$ keV & & & \\[0.1cm]
    & $\sigma$ & 0.1 keV & & & \\[0.1cm]
    & Flux$_{\rm unabsorb, gauss}$ & $6.6_{-4.2}^{+7.0} \times 10^{-16}$ ergs cm$^{-2}$ s$^{-1}$ & & & \\[0.1cm] \hline
\end{tabular}

\end{table*}

\subsubsection{4.5 keV feature?}
The apparent emission feature at $\sim 4.5$ keV could not be explained by any of the above models. There are 6 photons between 4.4 - 4.7 keV, corresponding to a count rate of $3.75 \times 10^{-5}$ counts/s. 
This is a small number of counts, and our $0\farcs5$ extraction region should only capture 60\% of the 4.5 keV point-spread function. Nevertheless, the line appears significant (as we show below), and we are unaware of any calibration issue that would generate a spurious emission line at this energy. We use the \texttt{flux} command in \textsc{XSPEC} to calculate the total flux in the interval $4.4 - 4.7$ keV and determine the predicted number of photons  using the exposure time ($= 1.59 \times 10^5 s$) and effective area of {\it Chandra} ACIS-S at $5$ keV ($= 400$cm$^2$). The model with two collisionally-ionized plasmas, for example, only predicts $1.23_{-1.0}^{+0.23}$ counts. Therefore we chose to further investigate the presence of an emission feature at this energy. This feature is consistent with the energy of the K$\alpha$ emission line of Ti, but the very high abundance of Ti needed ($\sim$1000) seems extremely improbable. We add a Gaussian with $\sigma = 0.1$ and line energy constrained between $4.4 - 4.7$ keV to the two-temperature plasma model. Adding the Gaussian greatly improves the fit ($\Delta \mathrm{cstat}_{2vapec}/\Delta \mathrm{d.o.f}_{2vapec} = 8.87/-2$), as shown in Fig.~\ref{fig:hen_vapec_gauss} and Table~\ref{table:spec_hen2-10}. We find $kT > 5$ keV for the hot plasma, $kT = 0.59_{-0.16}^{+0.20}$ keV for the cold plasma and the (linked) abundances of O, Ne, Mg, Si \& S to be $11_{-7}^{+46}$. From the C-statistic value ($=90.95$) and the number of parameters ($=8$), we can calculate the AICc value to be $108.44$. Comparing this to the AICc value of our two temperature model without the Gaussian ($=112.59$), we see that adding a Gaussian at $\sim 4.5$ keV increases the likelihood that the fit explains the data by a factor of $\sim 8$. If we treat $\Delta$C-statistic as a $\Delta \chi^2$-statistic (since both are logarithms of likelihood ratios), using the $\Delta \mathrm{cstat}_{2vapec}$ \& $\Delta \mathrm{d.o.f}$ observed, we have a 1.2\% probability of getting this statistical improvement by chance. Adding a Gaussian to the partially ionized hot plasma model gives similar results, with $\Delta$cstat $ = 10.1$ while decreasing the degrees of freedom by 2.

The potential presence of this line raises a critical question: are any similar lines observed in other SNRs? A faint emission line of Ti has been observed in Tycho's SNR at $4.7$ keV \citep{miceli2015}, but the 3.9 keV Ca He$\alpha$ line (undetectable in our data) is at least 240 times stronger (measured by equivalent width) in that SNR. An alternative scenario is that the line could be the $\sim$3.9 keV Ca line, blueshifted by $\sim$10\% of c. This would require that the Ca-emitting part of the remnant is moving at extreme speeds unusual in SNRs, and uniformly in our direction ---  both of which seem unlikely.  We conclude that the nature of this candidate line remains unclear.

\subsubsection{Nearby X-ray emission in Henize 2-10}

\begin{figure}
\includegraphics[width=\columnwidth]{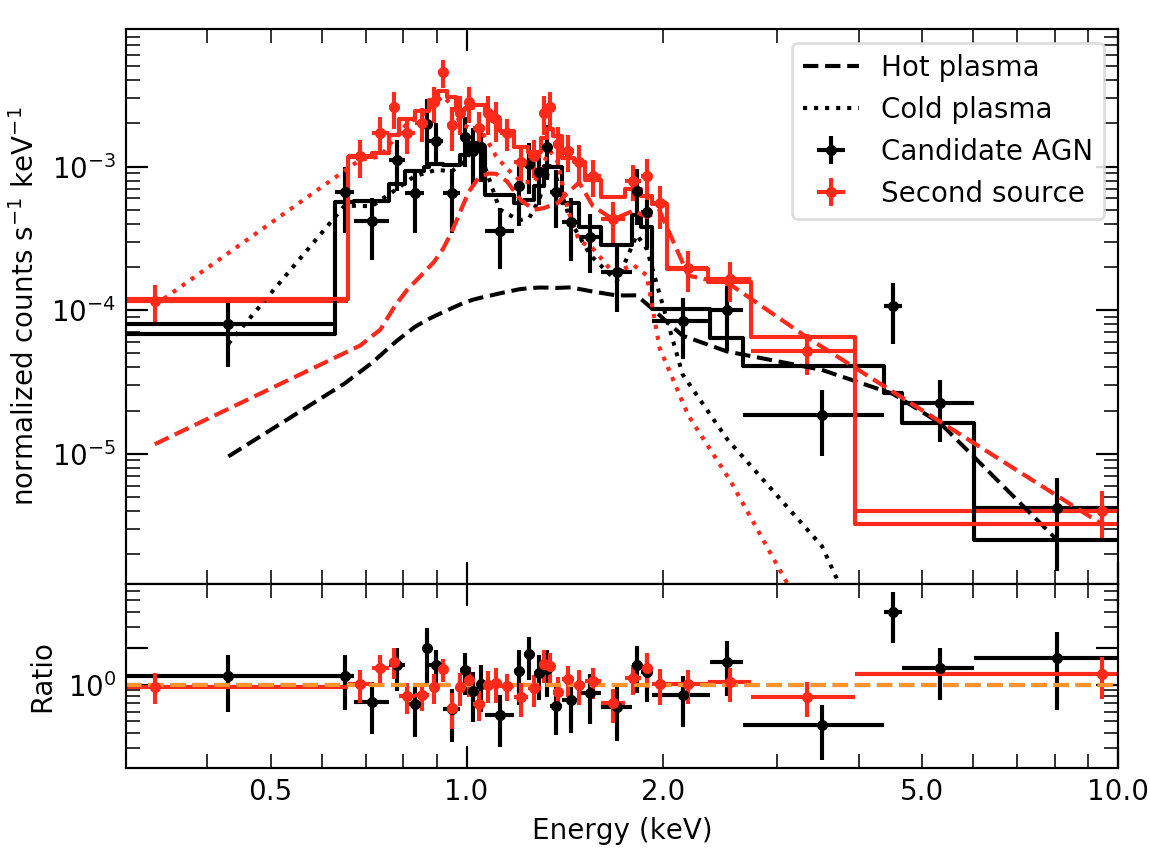}
\caption{ Comparison between the spectra of the two X-ray sources in Henize 2-10. The dashed and the dotted lines show the contribution of hot and cold plasma respectively to the total X-ray flux. We find that the X-ray spectrum of the candidate AGN is slightly harder, and has a higher required metallicity, than the other source. We also note that the second source in Henize 2-10 does not show any excess emission around 4.5 keV.}
\label{fig:henize_total}
\end{figure}

We also analyse the X-ray spectrum of the diffuse X-ray emission $1\arcsec$ east of the candidate AGN in Henize 2-10 (shown in green circle in Fig.~\ref{fig:source_bkg}), for comparison. We find that the X-ray spectrum of this eastern source is similar to our candidate AGN, with strong O, Ne, Mg and Si K$\alpha$ emission lines. Therefore, we use our two-temperature hot plasma model to explain the X-ray emission from this source. Fitting this model to its X-ray spectrum gives $N_H = (7 \pm 5) \times 10^{21}$ cm$^{-2}$, $kT_{\rm hot} = 1.9_{-0.6}^{+1.1}$ keV, $kT_{\rm cool} = 0.30_{-0.12}^{+0.33}$ keV, and the relative solar abundance of O, Ne, Mg \& Si to be $3_{-2}^{+4}$, i.e., only marginally super-solar. Fig.~\ref{fig:henize_total} compares the spectra of these two X-ray sources in Henize 2-10. From Fig.~\ref{fig:henize_total}, and the values of the best-fitting parameters, we see that though these sources have broadly similar spectra, the X-ray emission from the candidate AGN in Henize 2-10 is both harder  and requires a higher abundance of O, Ne, Mg \& Si compared to the diffuse emission. These differences help us investigate the nature of the differences between the two X-ray sources.

We note first that the contribution from this diffuse source to the spectrum of the candidate AGN is minimal.
From Fig.~\ref{fig:source_bkg}, we see that the two sources are separated by $> 1\arcsec$. A $1\arcsec$ radius encloses 84 (91)\% of the point-spread function for photons of 6.5 (1.5) keV. Some particular points of interest in the candidate AGN's spectrum --- the low-energy emission lines, the hard component, and the 4.5 keV possible line --- are all stronger in the candidate AGN than the diffuse eastern source, clearly indicating that none of them are produced by contamination from the eastern source. 

The similarities and differences between the two spectra suggest that the candidate AGN is likely produced by a single young SN, while the diffuse eastern source is produced by multiple, older SNRs. The eastern source's larger spatial extent would thus arise from multiple sources, while the candidate AGN is consistent with a point source. The higher-temperature spectrum of the candidate AGN suggests a faster expansion speed, typical of younger SNRs. Similarly, the higher metallicity is consistent with less mixing of ejecta with the circumstellar and interstellar medium.  This picture is consistent with the VLBI size constraints upon the radio source, which must be several times larger than 0.1 pc in size \citep{Ulvestad07}, but not much larger than 1.3 pc in size \citep[][cf. their resolving a larger SNR to the NE with VLBI]{Reines12}. In this picture, the candidate AGN is actually a young SNR that is less than 100 years old and is extremely luminous in the radio and X-ray, 4 and 3 times brighter than Cas A in these bands \citep[e.g.][]{Ulvestad07,Ou18}.

\subsubsection{Variability of Henize 2-10}
\citet{reines2016} identified evidence for (likely) periodic variability in the X-ray light curve of Henize 2-10.  This evidence does not come from a signal in a standard periodicity search (e.g.\ a power spectrum). Instead, a sinusoidal model (with period 33.5 ks) was statistically favoured compared to a constant rate model (at the 3.8$\sigma$ level), as measured with the F-test (typically used to compare spectral models). In addition, the amplitude of the fitted sinusoid was non-zero at 3.9$\sigma$ significance. Should significant variability on short timescales be confirmed, it would require a source of X-ray emission other than a SNR. For instance, a slow-spinning X-ray luminous magnetar within the supernova remnant, similar to the one in the SNR RCW 103 \citep{deluca2006,Rea16,D'Ai16} could explain such variability.

However, the methods used to verify variability by \citet{reines2016} do not appear to account for the number of trials used to search for the correct period, phase, and amplitude of the sinusoidal fit. Thus they may be less statistically significant than claimed.  The F-test is also designed to be used with Gaussian, not Gehrels, errors.  We therefore investigated the evidence for variability using several methods, and performed Monte Carlo testing to directly measure the statistical significance of nonstandard tests for  variability. For all timing analyses, we first barycenter the events file using the position of the target source.

Our first variability check uses the Gregory-Loredo variability test algorithm implemented in the \texttt{glvary} tool of CIAO\citep{Gregory92}\footnote{http://cxc.harvard.edu/csc/why/gregory\_loredo.html}. This algorithm was designed to search for periodic, variable signals, such as the one suggested here, as well as nonperiodic variabiliby.  Running this tool on the unbinned event file returns \textsc{VAR\_INDEX} $=0$, the lowest variability index, indicating no evidence for variability in the X-ray emission. 

We also calculate the Leahy normalised power spectral density (PSD) \citep{Leahy1983} to check the significance of the reported period, using the \texttt{powspec} tool in FTOOLS. We use the minimum time resolution of ACIS, 3.14101s, to construct background subtracted light curves to calculate the power-spectra. The PSD values at $P = 29.4$ ks and $P = 34.3$ ks are 7.68 and 10.44 respectively. We find that there are 173 out of the 51648 frequencies in our spectra that have PSD values greater than or equal to 10.44 (and 700 values greater than 7.68). Thus, the PSD value at the frequency reported by \citet{reines2016} does not appear to be significant. This is in agreement with \citet{reines2016}, who found a less than 2$\sigma$ significance for their claimed signal from a power spectral analysis. In sum, we do not see any evidence for this potential periodicity in these standard statistical tests.

We performed Monte-Carlo simulations to understand the significance of the difference in $\chi^2$ between the constant and sinusoidal fits, as well as the ratio of the amplitude of the sinusoidal model to its error, reported by \citet{reines2016}. We first binned the data using a bin time of 5 ks \citep[following][]{reines2016} and fit the data using a constant and sine models. We used Gehrels errors for the fitting. The best fitting constant model gives $4.7 \pm 0.6$ counts per bin and $\chi^2 = 19.25$ for 32 degrees of freedom, similar to \citet{reines2016} who reported $\chi^2/\nu = 18.4/32$. The sinusoidal model gives best fitting values of the amplitude, $A = 2.5 \pm 0.9$ counts/bin, and period, $P=(3.2 \pm 0.1) \times 10^4$ s, with  $\chi^2/\mathrm{d.o.f} = 11.39/29$ for the best fitting sine model. This is slightly larger than the fit value from \citet{reines2016}, $\chi^2/\nu$=9.2/29; the difference might be due to different background subtraction.

We then simulated $10^4$ realisations of 180 photons uniformly emitted over the observed time of 160 ks, and binned these realisations using the same bin time of 5 ks. We fit each of these realisations with constant and sinusoidal models using the {\sc Python SciPy}  \texttt{curve\_fit} function \citep{scipy2001}. To ensure that we identify the best fitting sine model, we explore ten different values of period and five values of amplitude as the initial parameters for fitting each realisation, and chose the best fitting model with the smallest $\chi^2$ values for each. We find that $14\%$ of the random realisations have $\chi^2 >= 19.25$ for the constant fit. We also check what fraction of our realisations show a ratio $A/\sigma_A >= 2.8$ (i.e., that the amplitude of the sinusoidal signal is detected at 2.8$\sigma$ confidence, as in the data). Finally, we check what fraction of the random realisations show an improvement in the sinusoidal fit compared to the constant fit that are stronger than that found in our date (i.e., $\chi^2_{\rm const} - \chi^2_{\rm sin} > 7.86$). We find that $3\%$ of our simulations satisfy both the conditions on the ratio of amplitudes and the difference in $\chi^2$. (The fraction of realisations satisfying either of the above conditions is also around $3\%$, as typically these conditions occur together.) We thus find that the probability of a non-variable light curve producing a signal of this strength is 3\%, which is moderately significant --- above 2$\sigma$, but not above 3$\sigma$. Although this possible signal is interesting, we do not feel that it is robust enough to rule out the non-variable SNR interpretation of this X-ray source.

\subsection{NGC 4178}
\label{sec: NGC4178}
\begin{figure}
\centering
\includegraphics[width=\columnwidth]{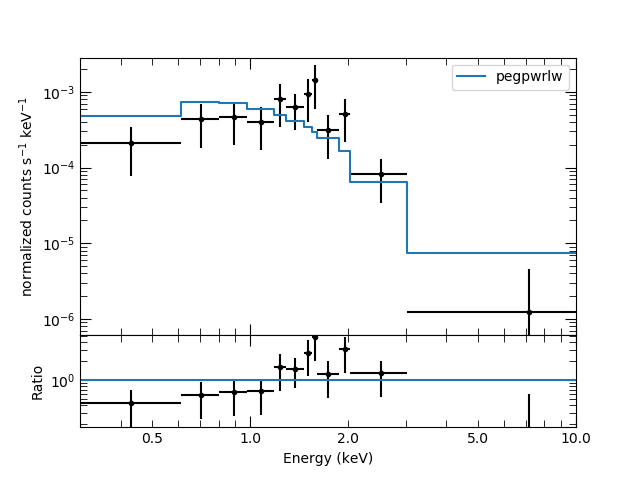}
\caption{X-ray spectrum of the candidate AGN in NGC 4178, fit with a partially-covered power-law model as in \citet{secrest2012}, but with finer binning; $N_H = 5 \times 10^{24}$cm$^{-2}$, covering fraction  $f = 0.99$, and  $\Gamma = 2.3$. The C-statistic for this model is $ 43.27$. 
The residual plot (below)
shows that this model predicts an excess of photons at both lower and higher energies, compared to the observed data.
} 
\label{fig:n4178_expected}
\end{figure}

\begin{table*}
\centering
\caption{Summary of spectral analysis results for the candidate AGN in  NGC 4178.}
\label{table:spec_ngc4178}
\begin{tabular}{c c c c c c c}
\hline
\multirow{2}{*}[0.0cm]{\bf Model} &	\multirow{2}{*}[0.0cm]{\bf Parameters} & \multirow{2}{*}[0cm]{\bf Parameter Values} &	\multirow{2}{*}[0cm]{$\mathbf{\Delta}${\bf cstat}$_{\mathbf{pl}}$/$\mathbf{\Delta}${\bf d.o.f}$_{\mathbf{pl}}$} &	\multirow{2}{*}[0cm]{\bf AICc value} &	{\bf log(CvM)} \\
 & & & & & {\bf (goodness)} \\
\hline
\multirow{4}{1in}[-0.1cm]{\textsc{pcfabs*pegpwrlw} \\(Expected)} &	$N_H$ &	$5 \times 10^{24}$ cm$^{-2}$ &	\multirow{4}{*}[-0.1cm]{$7.98/2$} & 	\multirow{4}{*}[-0.1cm]{$45.41$} &	\multirow{4}{0.5in}[-0.1cm]{\centering $-3.60$\\$(99\%)$} \\[0.1cm]
	& $f$ & $0.99$ & & & \\[0.1cm]
	& $\Gamma$ &	$2.3$ &	&	& 	\\[0.1cm]
    & Flux$_{\rm unabsorb}$ & $8.4_{-2.2}^{+2.6} \times 10^{-13}$ ergs cm$^{-2}$ s$^{-1}$ & & & \\[0.1cm] \hline
\multirow{3}{1in}[-0.1cm]{\textsc{tbabs*pegpwrlw}} &	$N_H$ &	$4.1_{-3.3}^{+5.1} \times 10^{21}$ cm$^{-2}$ &	\multirow{3}{*}[-0.1cm]{$0.0/0$} & 	\multirow{3}{*}[-0.1cm]{$42.18$} &	\multirow{3}{0.5in}[-0.1cm]{\centering $-5.14$\\$(99 \%)$} \\[0.1cm]
	& $\Gamma$ &	$2.9_{-1.0}^{+1.4}$ &	&	& 	\\[0.1cm]
    & Flux$_{\rm unabsorb}$ & $1.8_{-1.0}^{+8.1} \times 10^{-14}$ ergs cm$^{-2}$ s$^{-1}$ & & & \\[0.1cm] \hline
\multirow{3}{1in}[-0.5cm]{\textsc{tbabs*(pegpwrlw+ \\gaussian+gaussian)}} &	$N_H$ &	$3.4_{-3.2}^{+10.7} \times 10^{21}$ cm$^{-2*}$ &	\multirow{3}{*}[-0.5cm]{$-15.02/-4$} & 	\multirow{3}{*}[-0.5cm]{$39.14$} &	\multirow{3}{0.5in}[-0.5cm]{\centering $-7.62$\\$(30 \%)$} \\[0.1cm]
	& $\Gamma$ &	$4_{-2}^{+10}$ &	&	& 	\\[0.1cm]
    & Flux$_{\rm pl, 0.2-10.0}$ & $1.8_{-1.5}^{+68} \times 10^{-14}$ ergs cm$^{-2}$ s$^{-1}$ & & & \\[0.1cm]
    & $E_1$  & $1.5 \pm 0.1$ keV & & & \\[0.1cm]
    & $E_2$ & $2.0 \pm 0.1$ keV & & & \\[0.1cm]
    & $\sigma_1, \sigma_2$ & $0.1$ keV & & & \\[0.1cm] \hline
\multirow{5}{1in}{\vspace{-0.3cm} \textsc{tbabs*vapec}} & $N_H$ &	$9.4_{-7.2}^{+27.6} \times 10^{20}$ cm$^{-2 \ *}$ &		\multirow{5}{*}{\vspace{-0.3cm} $-8.33/-1$} & \multirow{5}{*}{\vspace{-0.3cm} $36.50$} &	\multirow{5}{0.5in}[-0.1cm]{\centering $-6.49$\\$(45\%)$} \\[0.1cm]
	& $kT$	& $1.4_{-0.4}^{+1.1}$ keV&	&	&	\\[0.1cm]
	& O, Ne, Mg, Si &	$761_{-750}^{+240\ **}$ & & & \\[0.1cm]
    & E.M.$^a$ & $10^{58} - 10^{61}$ cm$^{-3}$& & & \\[0.1cm] \hline
\end{tabular} \\
\justify
Note: The model given by \citet{secrest2012} (line 1) cannot explain the observed results. Comparing the $\Delta$cstat/$\Delta$d.o.f, AICc and the CvM statistic values, it is more probable that NGC 4178 is a supernova remnant.\newline
$^{*}$ Lower error bound reaches the lower hard limit. \newline
$^{**}$ Upper error bound is more than the upper hard limit for the parameter. \newline
$^a$ E.M. stands for Emission Measure  = $\int n_e n_H dV$.
\end{table*}

We analyse the 36 photons from the candidate AGN in NGC 4178 using similar methods. \citet{secrest2012} uses a partially absorbed power-law  model with $N_H = 5 \times 10^{24}$cm$^{-2}$, a covering fraction $f = 0.99$, and  $\Gamma = 2.3_{-0.5}^{+0.6}$ to explain the observed X-ray flux and the hardness ratio. We fit this model to the minimally binned X-ray spectra in Fig.~\ref{fig:n4178_expected}. This model does not match the strong emission between 1.8 and 2.0 keV, and over-predicts emission above 2 keV and below 1 keV. Inspection of the residuals suggests the possibility of Si emission lines between 1.8 and 2 keV, and Mg lines around 1.4 keV, motivating a thermal plasma model.

\begin{figure}
\centering
\includegraphics[width=\columnwidth]{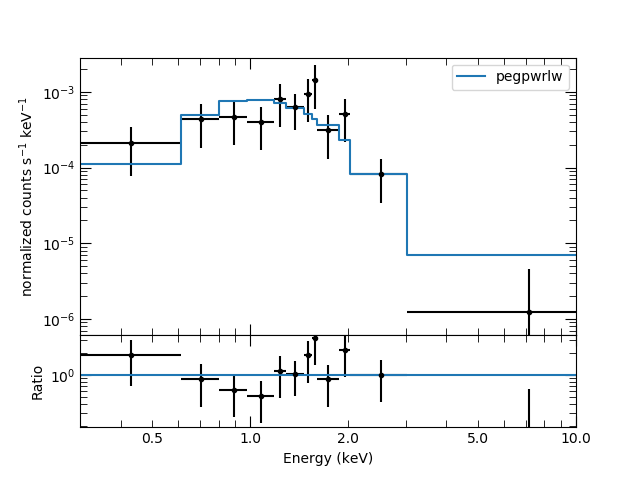}
\caption{X-ray spectrum of the AGN candidate in NGC~4178, here modelled by a
power-law, with $N_H = 4_{-3}^{+5} \times 10^{21}$ and $\Gamma = 2.9_{-1.9}^{+1.4}$. C-statistic $= 35.3$. Though this model is a better fit than the one in \citet{secrest2012} and Fig.~\ref{fig:n4178_expected}, it does not model the potential emission lines.}
\label{fig:n4178_pegpw}
\end{figure}

\begin{figure}
\centering
\includegraphics[width=\columnwidth]{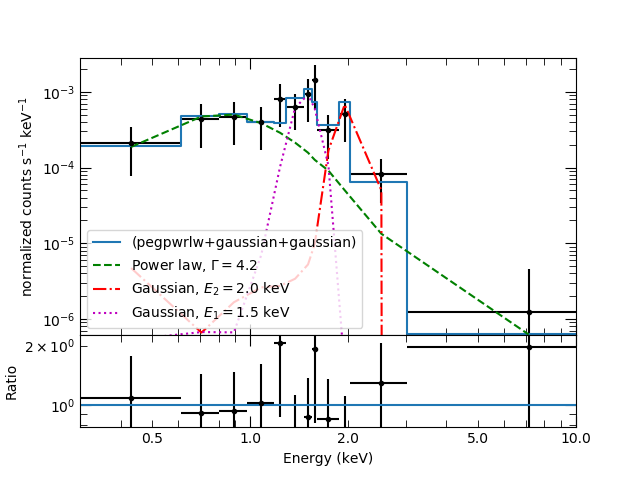}
\caption{X-ray spectrum of the candidate AGN in NGC 4178, modelled by a simple power-law with two Gaussian lines, with $N_H = 3.4_{-2.2}^{+10.7} \times 10^{21}$ cm$^{-2}$, $\Gamma = 4_{-2}^{+10.7}$. These Gaussian lines indicate excess emission at $1.5 \pm 0.1$ and $2.0 \pm 0.1$ keV, which coincide with the K$\beta$ emission of Mg XII (= 1.473 keV) and Si XIV (= 2.004 keV). C-statistic = $20.3$.}
\end{figure}

We also check a simpler power-law fit (\textsc{tbabs*pegpwrlw}) to the NGC 4178 X-ray spectrum, finding $N_H = 4.1_{-3.3}^{+5.1} \times 10^{21}$ cm$^{-2}$ and $\Gamma = 2.9_{-1.0}^{+1.4}$. From Figs.s~\ref{fig:n4178_expected} and ~\ref{fig:n4178_pegpw}, we see that neither power-law model fits the observed emission lines or rapid count rate decrease above 2 keV. The ``goodness'' simulations for both these models show that a large fraction ($\sim 99\%$) of the simulated realizations have a lower CvM statistic than the data, suggesting that these fits are poor. The large magnitude of the photon index for a power-law fit also suggests a softer thermal origin of the X-ray emission.

We check the significance of excess emission near 1.5 keV and 2.0 keV by adding Gaussian emission lines to the simple power-law, as in the previous case. These lines could correspond to the K$\beta$ emission lines of Mg and Si. Gaussian lines with $\sigma = 0.1$ produce good fits at $E_1 = 1.5 \pm 0.1$ keV and $E_2 = 2.0 \pm 0.1$ keV. These Gaussian lines reduce the C-statistic of the best fitting model by $\sim 15$ while decreasing the d.o.f by 4. From the AICc value of this model (39.14), we find that this model is $\sim 5$ times better than the simple power-law model. Running the ``goodness" test with this model shows that $\sim 30\%$ of the test models had CvM values smaller than the best fitting model, indicating that the model is a reasonable representation of the data.

\begin{figure}
\centering
\includegraphics[width=\columnwidth]{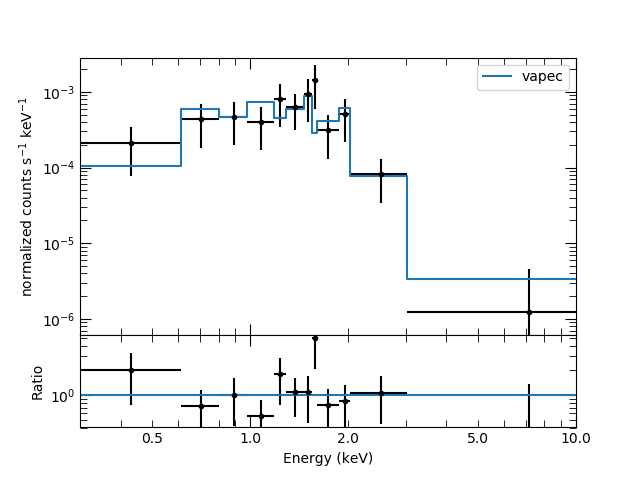}
\caption{X-ray spectrum of the AGN candidate in NGC~4178, here modelled by a collisionally ionized hot plasma model, with $N_H = 9.4_{-7.2}^{+27.6 \times 10^{20}}$ cm$^{-2}$, \& $kT = 1.4_{-0.4}^{1.1}$ keV. C-statistic $= 26.9$. This model is able to fit the narrow emission features along with the broad characteristics of the spectrum.  The AICc test indicates that the likelihood of this hot plasma model is $\sim 17$ times larger than that of a simple power-law model.}
\label{fig:n4178_vapec}
\end{figure}

To explain the emission features, we try fitting a single temperature \textsc{tbabs*vapec} model. Since we have very low photon counts, we link the relative abundances of O, Ne, Mg \& Si, reducing the number of parameters. Fitting this model to the observed spectrum (Fig.~\ref{fig:n4178_vapec}) gives $N_H = 9_{-7}^{+28} \times 10^{20}$ cm$^{-2}$, $kT = 1.4_{-0.4}^{+1.1}$ keV, and super-solar abundances of O,  Ne, Mg and Si ($> 11$). With only one additional parameter compared to the absorbed power-law fit, the C-statistic reduces by $8.3$. Thus comparing the AICc values, the hot plasma model is about 17 times more probable to explain the observed spectrum, compared to the simple power-law model. The lack of photons above 2.1 keV, matching the single temperature plasma model predictions, indicates that this model is sufficient to explain the observed data. The super-solar abundances of Mg and Si indicate that the X-rays from the candidate AGN in NGC 4178 can also be better explained as emission from a SNR than an AGN.

\section{Summary and Conclusions}
We have used the moderate spectral resolution of {\it Chandra}, and the strength of emission lines in the thermal emission from SNRs, to discriminate between power-law and thermal plasma emission models in low-count X-ray spectra. Through our analysis, we argue that the X-ray sources identified as  candidate AGN in Henize 2-10 and NGC 4178 are more likely to be supernova remnants than actual AGN. 

The recent analysis of \citet{cresci2017}  argued that the optical emission lines from the AGN candidate in Hen 2-10 indicate ionization by star formation rather than an AGN. In addition, the revision of the X-ray luminosity of the candidate AGN by \citet{reines2016} reduces its X-ray/radio flux ratio into the regime of SNRs \citep{cresci2017}. 

Thus, we find that the published evidence from optical, X-ray, and radio emission does not make a compelling argument for an AGN in Henize 2-10.  However, Reines et al. (in prep.; priv. comm.) argue that their newly obtained HST/STIS spectra of the AGN candidate favour LINER-like line ratios.

We find that the X-ray source identified by \citet{secrest2012} as a candidate highly-obscured AGN does not resemble the X-ray spectrum of an obscured AGN, and instead can be well described by models for SNRs.  
The only remaining evidence for an AGN in NGC 4178 is the strength of the [Ne V] line \citep{satyapal2009}, suggested to be produced by an obscured AGN. Given the lack of optical \citep{secrest2013} and X-ray evidence for an AGN in this galaxy, we suggest that radiative transfer simulations be performed to determine if the [Ne V] line might be produced by a young stellar population as well.

Although there is no question that small AGN exist in many dwarf galaxies, our work demonstrates that X-ray emission from SNRs can be a confounding factor in searches for low-luminosity AGN.

Standard hardness ratios, as well as more sophisticated quantile analyses \citep{Hong04}, do not take advantage of the capability of X-ray CCD spectra to resolve clear X-ray lines. We are now devising and testing general hardness ratios that can be used by X-ray CCD instruments to discriminate between strongly line-dominated and continuum-dominated spectra, even with low numbers of X-ray photon counts. Immediate applications include identifying Fe-K lines (e.g. from highly absorbed hard X-ray sources such as IGR J16318-4848, \citealt{Walter03}), Fe-L lines (around 1 keV) from chromospherically active stars with soft X-ray spectra, and Si lines from supernova remnants. 

Future X-ray and radio instruments will also permit more powerful analyses. For objects such as Henize 2-10, for instance, the {\it Lynx} mission \citep{Lynx18}, with quantum calorimeter spectral resolution, high effective area, and {\it Chandra}-like angular resolution, can provide conclusive answers. Similarly, the ngVLA \citep{Murphy18}, with higher effective area and longer baselines than the Jansky VLA, would resolve radio sources such as this, enabling firm identification of low-luminosity low-mass AGN \citep{Plotkin18,Nyland18}.

\section*{Acknowledgements}
We thank A. Reines for discussions, particularly for stressing the importance of addressing the suggested variability, and the referee for a thoughtful report. PRH thanks E. Rosolowsky for discussions about the mid-infrared spectrum of NGC 4178.  COH and GRS are supported by NSERC Discovery Grants (NSERC RGPIN-2016-04602 and RGPIN-2016-06569, respectively), and COH also by a Discovery Accelerator Supplement.

\bibliographystyle{mnras}
\bibliography{henize2-10_ngc4178}

\bsp	
\label{lastpage}
\end{document}